\documentclass[oneside,reqno,english]{amsart}
\usepackage{amsmath}
\usepackage{graphicx}
 \usepackage[colorlinks=true]{hyperref}
\hypersetup{urlcolor=blue, citecolor=red}

\def\currentvolume{X}
 \def\currentissue{X}
  \def\currentyear{200X}
   \def\currentmonth{XX}
    \def\ppages{X--XX}
     \def\DOI{10.3934/xx.xx.xx.xx}

\theoremstyle{definition}

\usepackage{bm,amssymb}						
\usepackage[pagewise]{lineno}

\title[Entropic property of the ES model with $\mathrm{Pr}<2/3$] 
      {On the entropic property of \\ the Ellipsoidal Statistical model\\ with the Prandtl number below 2/3}

\author[Shigeru Takata, Masanari Hattori and Takumu Miyauchi]{}


 \email{takata.shigeru.4a@kyoto-u.ac.jp}
 \email{hattori.masanari.4r@kyoto-u.ac.jp}
 \email{miyauchi.takumu.23e@st.kyoto-u.ac.jp}

\thanks{The present work is supported in part by KAKENHI from JSPS (No. 17K18840)}

\thanks{$^*$ Corresponding author: Shigeru TAKATA}

\begin{document}
\maketitle

\centerline{\scshape Shigeru Takata$^*$ and Masanari Hattori}
\medskip
{\footnotesize
 \centerline{Department of Aeronautics and Astronautics 
\& Advanced Engineering Research Center,}
   \centerline{Kyoto University, Kyoto 615-8540, Japan}
} 

\medskip

\centerline{\scshape Takumu Miyauchi}
\medskip
{\footnotesize
 \centerline{Department of Aeronautics and Astronautics, Kyoto University, Kyoto 615-8540, Japan}
}

\bigskip

 \centerline{(Communicated by *****)}

\begin{abstract}
Entropic property of the Ellipsoidal Statistical model with the Prandtl number Pr below 2/3 is discussed. 
Although 2/3 is the lower bound of Pr for the H theorem to hold unconditionally,
it is shown that the theorem still holds even for $\mathrm{Pr}<2/3$, provided that 
anisotropy of stress tensor satisfies a certain criterion.
The practical tolerance of that criterion is assessed numerically 
by the strong normal shock wave and the Couette flow problems.
A couple of moving plate tests are also conducted.
\end{abstract}

\section{Introduction}

The Ellipsoidal Statistical (ES) model is a relaxation-type kinetic equation proposed by Holway \cite{H66}, 
which involves the celebrated Bhatnagar--Gross--Krook (BGK) model \cite{BGK54} as a special case.
It has a simple structure as the other kinetic models 
but still satisfies the H theorem and reproduces a realistic value of the Prandtl number.
Since the proof of its H theorem in Ref.~\cite{ATPP00},
it has been attracting many researchers as an alternative to the BGK model,
e.g., \cite{B08,GT11,GMS11,MWRZ13,KPP16}.

In the case of monatomic gases, that proof was made for the Prandtl number not less than 2/3.
In some practical applications, however, such a restriction might affect our choice of the Prandtl number. 
Indeed, for the hard-sphere gas (one of the most representative models in the kinetic theory) the Prandtl number is known to be  
$0.660694$ \cite{S07}, which is slightly lower than $2/3$.%
\footnote{In the case of the dense gas, 
the deviation from $2/3$ can be enhanced to be down to about $86\%$ 
of the value at the dilute gas limit,
according to the Enskog theory; 
see, e.g., J. O. Hirschfelder, C. F. Curtiss and R. B. Bird, \textit{Molecular Theory of Gases and Liquids}, John Wiley \& Sons, New York, 1964, Sec.~9.3.} 
In Refs.~\cite{TFA10,FTAK11},
we were also in such an embarrassing situation 
that the Prandtl number of nitrogen gas is slightly below the corresponding bound for polyatomic gases
and made a compromise (by around 10 \%) in setting the value of the Prandtl number
in numerical computations.

The above lower bound $2/3$ is, however, the condition that the H theorem holds unconditionally.
There is still a room for that the H theorem holds even for the Prandtl number below that bound under some restriction. 
In the present paper, we are going to study this issue
and provide a critical condition for the H theorem
to be valid for the Prandtl number below 2/3.
The paper is organized as follows. First, the ES model is briefly reviewed in Sec.~\ref{sec:ESH}. Then, the derivation of the concerned condition is delivered in Sec.~\ref{sec:Derivation}, which is followed by some practical assessment tests in Sec.~\ref{sec:tests}. The paper is concluded in Sec.~\ref{sec:conclusion}.

\section{ES model and H theorem}\label{sec:ESH}

Let us denote by $f$ the velocity distribution function (VDF) of gas molecules,
which is a function of time $t$, space position $X_i$, and molecular velocity $\xi_i$.
Then, the collision term of the ES model $Q_\mathrm{ES}(f)$ for monatomic gases reads 
\cite{ATPP00}
\begin{subequations}
\begin{align}
Q_\mathrm{ES}(f)=& A_c(T)\rho(\mathcal{G}[f]-f), \\
& \mathcal{G}[f]=\frac{\rho}{\det(2\pi\mathcal{T})}
  \exp\Big(-\frac{1}{2}(\mathcal{T}^{-1})_{ij}(\xi_i-v_i)(\xi_j-v_j)\Big), \label{eq:Gauss}\\
& \rho = \int f d\bm\xi, \quad v_i=\frac{1}{\rho}\int \xi_i f d\bm\xi, \quad
  \mathcal{T}_{ij}=(1-\nu)RT\delta_{ij}+\nu\,\Theta_{ij}, \label{eq:CalT}\\
& \Theta_{ij}=\frac{1}{\rho}\int (\xi_i-v_i)(\xi_j-v_j)fd\bm\xi, 
  \quad T=\frac{1}{3R} \Theta_{ii},\label{eq:Theta}
\end{align}
\end{subequations}
where $\rho$, $v_i$, and $T$ are respectively the density, flow velocity, and temperature of the gas; $R$ is the specific gas constant (the Boltzmann constant divided by the mass of a molecule); $\delta_{ij}$ is the Kronecker delta;
$A_c(T)(>0)$ is a positive function of $T$ such that $A_c(T)\rho$ is the collision frequency of the gas molecules.
Note that $\mathcal{T}$ (or $\mathcal{T}_{ij}$) is a $3\times3$ matrix, 
that $\Theta_{ij}$ (or $\Theta$) is the stress tensor divided by the density,
and that the summation convention applies 
in the definitions of $\mathcal{G}$ and $T$ and throughout the present paper. 
The $\nu$ is an adjustable parameter to make the Prandtl number $\mathrm{Pr}$ 
a realistic value, 
and the H theorem is proved to hold unconditionally in the range $-1/2\le\nu<1$ \cite{ATPP00}.
Because of the relation $\mathrm{Pr}=1/(1-\nu)$ (or $\nu=1-1/\mathrm{Pr}$),
this range is identical to $2/3\le\mathrm{Pr}<\infty$.

For our purpose, the relevant steps in the proof in Ref.~\cite{ATPP00} 
are to show that (i) $\mathcal{T}$ is positive definite and 
that (ii) the inequality $\det\mathcal{T}\ge\det\Theta$ holds,
where the equality holds only when $\Theta$ is a scalar multiple of the identity matrix 
(see the Appendix~\ref{sec:app}).  
If (i) and (ii) remain valid even for $\nu<-1/2$ (or $\mathrm{Pr}<2/3$) under some condition,
the H theorem remains valid as well under that condition.
In the present paper, we are concerned with the case
$-1\le\nu<-1/2$, which corresponds to the case $1/2\le\mathrm{Pr}<2/3$.

\section{Derivation of condition}\label{sec:Derivation}

Thanks to its definition, $\Theta$ is a symmetric positive definite matrix
and thus has positive real eigenvalues $\lambda_1$,  $\lambda_2$, 
and $\lambda_3$, as far as the VDF is non-negative.
Hence, without loss of generality, we may assume that $0<\lambda_1\le\lambda_2\le\lambda_3$ and accordingly 
may put $\lambda_1=(1-\epsilon)\lambda$, $\lambda_2=(1-\theta\epsilon)\lambda$,
and $\lambda_3=\lambda$, where $0\le\epsilon<1$, $0\le\theta\le 1$, and $\lambda>0$.
Since by definition $3RT$ is nothing else than the trace of $\Theta$, it is related to the eigenvalues
as $3RT=\lambda_1+\lambda_2+\lambda_3$ [or $3RT=(3-\epsilon(1+\theta))\lambda$].
Note that $\epsilon$ indicates the degree of deviation of the minimum 
from the maximum eigenvalue and that $\theta$ is the ratio of the difference between the largest two eigenvalues to that between the minimum and the maximum eigenvalue.


\subsection{On the condition for $\mathcal{T}$ to be positive definite}
In the frame of the mutually orthogonal eigenvectors of $\Theta$,
the $\mathcal{T}$ as well as $\Theta$ is diagonalized:
\begin{align*}
\mathcal{T}
&=\nu\Theta+(1-\nu)RT \mathrm{Id} \notag \\
&=\lambda\,
  \Biggl( \begin{array}{ccc}
  1-\epsilon\nu-\frac{\epsilon(1-\nu)(1+\theta)}{3} & 0 & 0 \\
  0 & 1-\epsilon\theta\nu-\frac{\epsilon(1-\nu)(1+\theta)}{3} & 0 \\
  0 & 0 & 1-\frac{\epsilon(1-\nu)(1+\theta)}{3} 
  \end{array} \Biggr ),
\end{align*} 
where $\mathrm{Id}$ is the $3\times3$ identity matrix.
Since the three diagonal components multiplied by $\lambda$ are eigenvalues of $\mathcal{T}$,
the condition for $\mathcal{T}$ to be positive definite is simply reduced to
\begin{equation}
1-(\epsilon/3)(1-\nu)(1+\theta)>0,\quad\mathrm{i.e.,}\quad
0\le\epsilon<\frac{3}{1-\nu}\frac{1}{1+\theta},
\label{cond_T}
\end{equation}
in the present parameter range $-1\le\nu<-1/2$.
Note that,
 since $0\leq\epsilon<1$ by construction,
the constraint \eqref{cond_T} is effective only when 
the right-hand side of the last inequality is less than unity, namely
%
\begin{equation}
\theta>\frac{3}{1-\nu}-1=\frac{2+\nu}{1-\nu}\equiv\theta_\mathcal{T}(\nu),
\label{theta_T}
\end{equation}
is fulfilled.
The above $\theta_\mathcal{T}$ varies monotonically from $1/2$ to $1$ 
when $\nu$ varies from $-1$ to $-1/2$.

\subsection{On the condition for $\det\mathcal{T}\ge\det\Theta$ to hold and the restriction on the equality}

After straightforward manipulations, the difference $D$ of $\det\mathcal{T}$ with $\det\Theta$ 
is written as
\begin{align}
D\equiv & \det\mathcal{T}-\det\Theta \notag \\
=& \frac13\lambda^3\epsilon^2(1-\nu)
   \Big\{
   (1+\nu)(\theta^2-\theta+1)\notag \\
&  -\frac19\epsilon(1+\theta)[2\nu+1+(1-\nu)\theta][(2\nu+1)\theta+1-\nu]
   \Big\}.
\end{align}
Here, we need to care the restriction 
 in (ii) of the last paragraph of Sec.~\ref{sec:ESH} that 
$D=0$ holds only when $\Theta$ is a scalar multiple of the identity matrix, 
which is identical to $\epsilon=0$ in the present notation.
Hence the condition for $D\ge0$ with this restriction
is reduced to that the inside of the curly brackets is positive, i.e.,
\begin{equation}
   (1+\nu)(\theta^2-\theta+1)
 > \frac19\epsilon(1+\theta)[2\nu+1+(1-\nu)\theta][(2\nu+1)\theta+1-\nu],
\end{equation}
for $\epsilon\ne0$ in the range $-1\le\nu<-1/2$.
The left-hand side is non-negative because $0\le\theta\le 1$, 
while the sign of the right-hand side depends on that of the factor 
$F(\theta;\nu)\equiv[2\nu+1+(1-\nu)\theta][(2\nu+1)\theta+1-\nu]$.
Let the smaller and larger zeros of $F$ with fixed $\nu$ be $\theta_a$ and $\theta_b$,
i.e., $\theta_a=-(2\nu+1)/(1-\nu)$ and $\theta_b=-(1-\nu)/(2\nu+1)$.
Then, it is readily seen that $0<\theta_a<1<\theta_b$
and that $F\ge0$ for $\theta_a\le\theta\le\theta_b$ and $F<0$ elsewhere.
Hence, we have the following condition:
\begin{equation}
\Biggl\{
\begin{array}{ll}
\displaystyle
 0\le\epsilon< 1, & \mbox{for }\ 0\le\theta\le\theta_a, \\
\displaystyle
 0\le \epsilon <
 \frac{9(1+\nu)(\theta^2-\theta+1)}{(1+\theta)F(\theta;\nu)}\equiv\mathcal{F}(\theta;\nu),
 & \mbox{for }\ \theta_a< \theta\le 1.
\end{array}
\Biggr .
\label{cond_D}
\end{equation}

\subsection{Combined condition}
We shall merge the conditions \eqref{cond_T} and \eqref{cond_D}
by taking account of the original range of $\epsilon$, i.e., $0\le\epsilon<1$.

Let us start with the observation that $\theta_\mathcal{T}-\theta_a=3(\nu+1)/(1-\nu)\ge0$, so that $\theta_a\le\theta_\mathcal{T}<1$ for $-1\le\nu<-1/2$.
As described before, the condition \eqref{cond_T} becomes effective only when $\theta>\theta_\mathcal{T}$.
Our first concern is the size relation between $3/[(1-\nu)(1+\theta)]$ in \eqref{cond_T} and $\mathcal{F}(\theta;\nu)$ in \eqref{cond_D} in the range $\theta_\mathcal{T}<\theta\le 1$.
Then, checking the size relation is reduced to checking the sign of the following quantity
\begin{equation*}
\Delta\equiv F(\theta;\nu)-3(1-\nu)(1+\nu)(\theta^2-\theta+1),
\end{equation*}
because all of $(1+\theta)$, $(1-\nu)$, and $F$ are positive.
Here, $\Delta\gtrless 0$ implies $3/[(1-\nu)(1+\theta)]\gtrless \mathcal{F}$.
The simple substitution of $F$ shows, after some manipulations, that
\begin{align}
\Delta
&=(1-\nu)(-\nu-2)\theta^2+(2\nu^2+2\nu+5)\theta
  +(1-\nu)(-\nu-2) \notag \\
&=[(1-\nu)\theta-(\nu+2)][-(\nu+2)\theta+(1-\nu)]\notag \\
&=-(1-\nu)(\nu+2)(\theta-\theta_\mathcal{T})(\theta-\frac{1}{\theta_\mathcal{T}}).
\end{align}
Remind that $\theta_\mathcal{T}<1$ for $-1\le \nu<-1/2$.
Thus $\Delta>0$, i.e., $\mathcal{F}<3/[(1-\nu)(1+\theta)]$ for $\theta_\mathcal{T}<\theta\le 1$.
Our next concern is the size relation between $\mathcal{F}$ and unity in the range $\theta_a<\theta\le\theta_\mathcal{T}$, because $3/[(1-\nu)(1+\theta)]$ is over or equal to unity in this range.
Then, because $\Delta\le0$ for $\theta_a<\theta\le\theta_\mathcal{T}$,
we have $1\le 3/[(1-\nu)(1+\theta)]\le \mathcal{F}$, 
so that there is no restriction on $\epsilon$ in the concerned range of $\theta$.

To summarize, we have arrived at the following statement of criterion.

Let $\lambda_1\leq\lambda_2\leq\lambda_3$
be positive real eigenvalues of the symmetric positive definite
matrix $\Theta$ defined by \eqref{eq:Theta}.
Let $\epsilon=1-\lambda_1/\lambda_3$
and $\theta=(1-\lambda_2/\lambda_3)/(1-\lambda_1/\lambda_3)$,
where $0\le\epsilon<1$ and $0\le\theta\le 1$ by construction.
Then, for $-1\leq\nu<-1/2$, which corresponds to $1/2\leq\mathrm{Pr}<2/3$, the following criterion must be fulfilled in order for the H theorem to remain valid:
%
%
\begin{equation}
\Biggl\{
\begin{array}{ll}
\displaystyle
 0\le\epsilon< 1, & \mbox{for }\displaystyle\ 0\le\theta\le\frac{2+\nu}{1-\nu}, \\
\displaystyle
 0\le \epsilon < \mathcal{F}(\theta;\nu),
 & \mbox{for }\displaystyle\ \frac{2+\nu}{1-\nu}< \theta\le 1,
\end{array}
\Biggr .
\label{CC}
\end{equation}
where $\mathcal{F}$ is the one defined by \eqref{cond_D}.

\section{Simple test problems}\label{sec:tests}

In order to see how much the derived condition \eqref{CC} is practically restrictive,
we shall carry out a couple of numerical tests 
such that the strong anisotropic behavior of the gas be expected.
One is the strong shock wave problem, the other is the time-dependent Couette flow problem. 
A couple of additional tests, which are simple but severe,
motivated by the result of the Couette flow will be conducted as well.

\subsection{Shock wave test}
\begin{figure}\centering
\begin{tabular}{cc}
\includegraphics[scale=0.25]{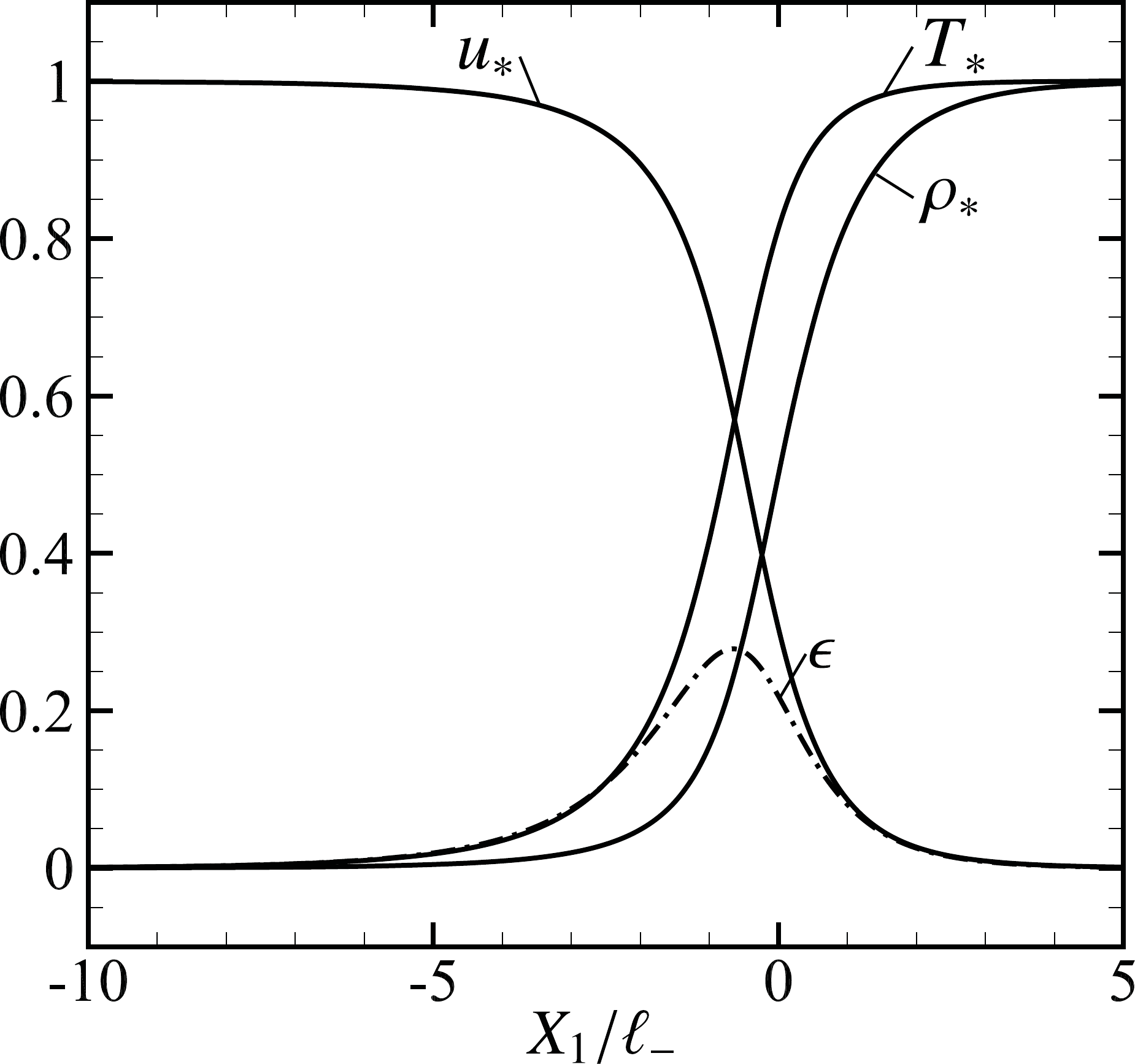} &
\includegraphics[scale=0.25]{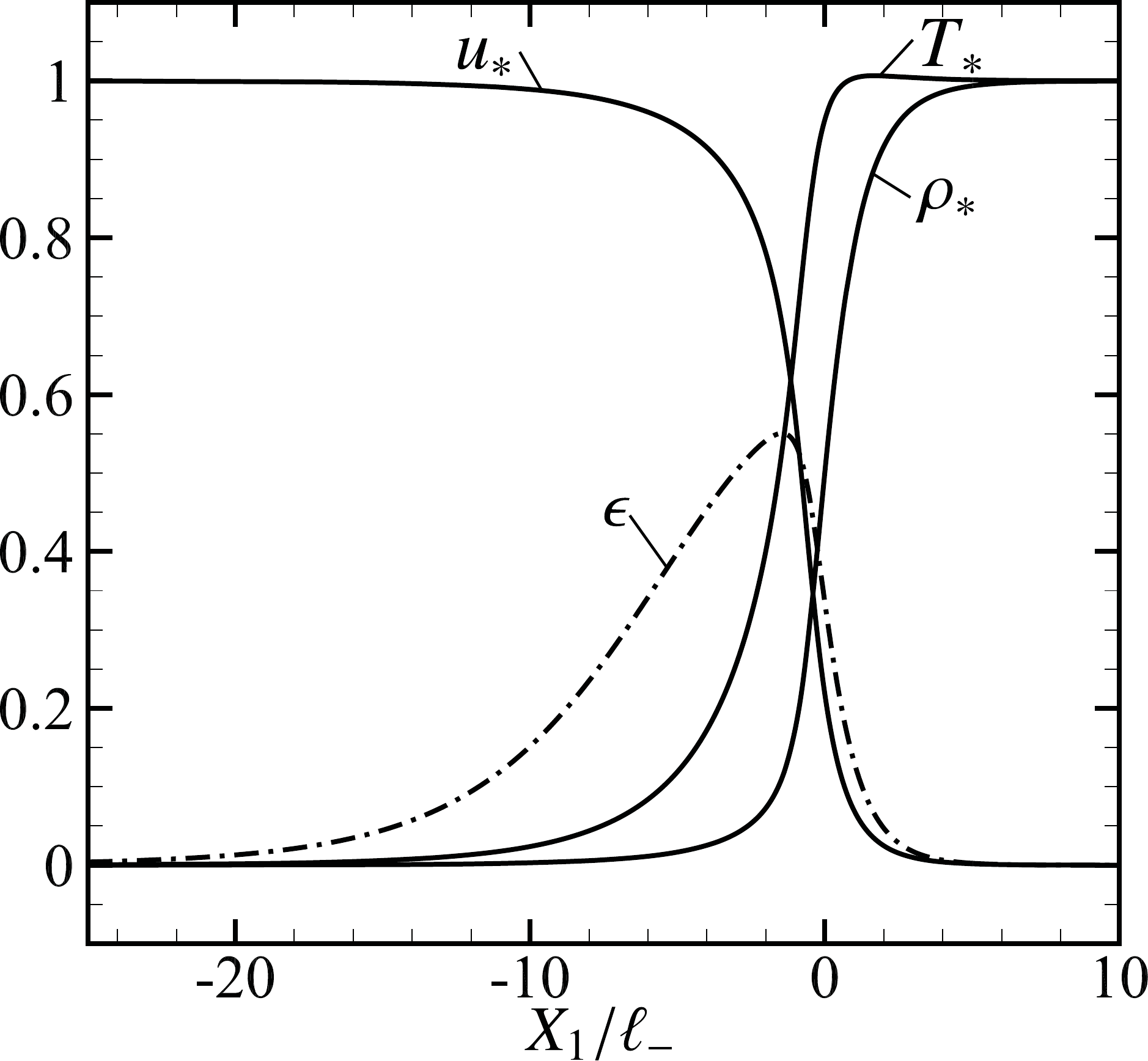} \\  
(a) $M_-=2$ & (b) $M_-=5$ \\ \\
\includegraphics[scale=0.25]{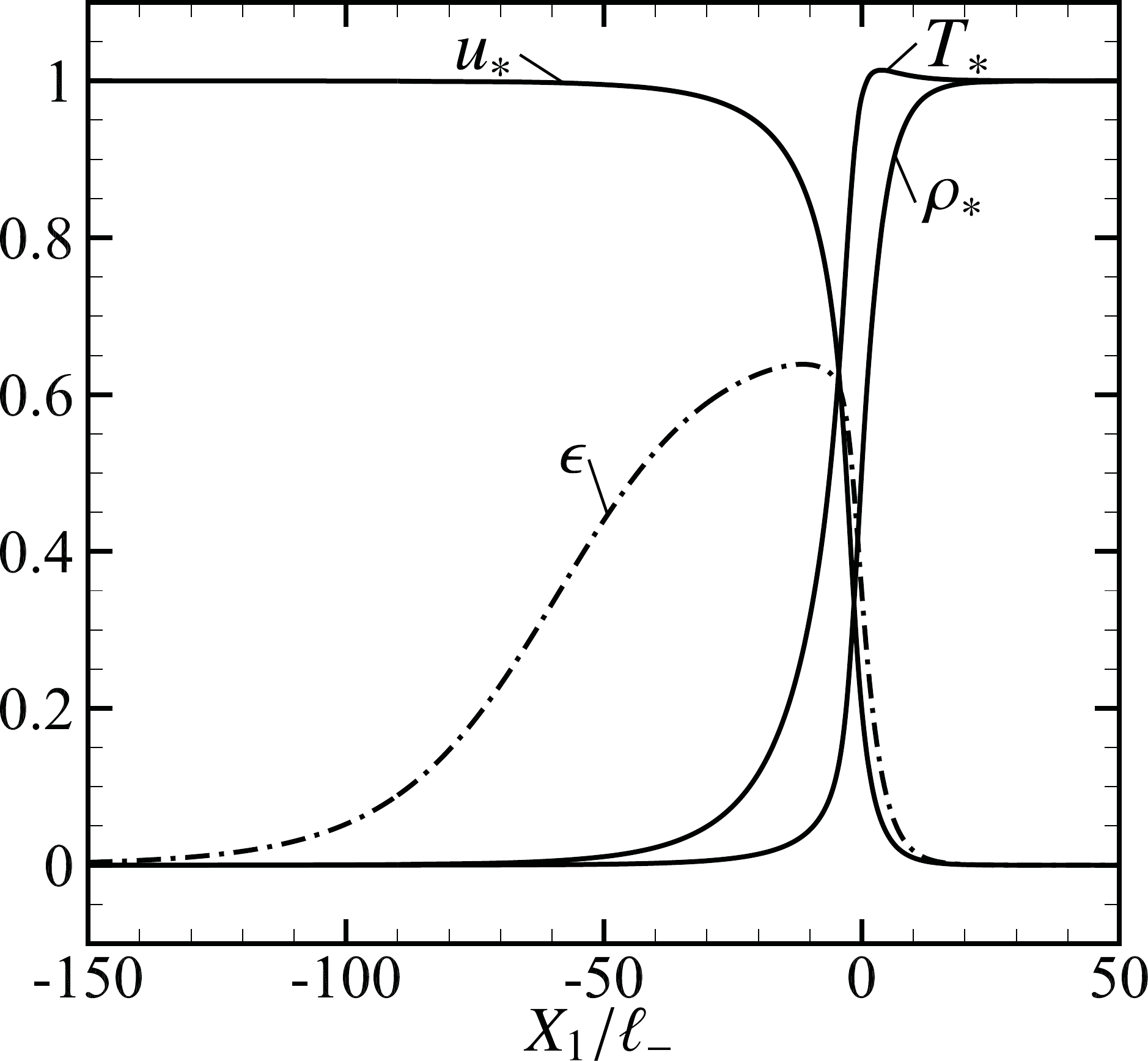} &
\includegraphics[scale=0.25]{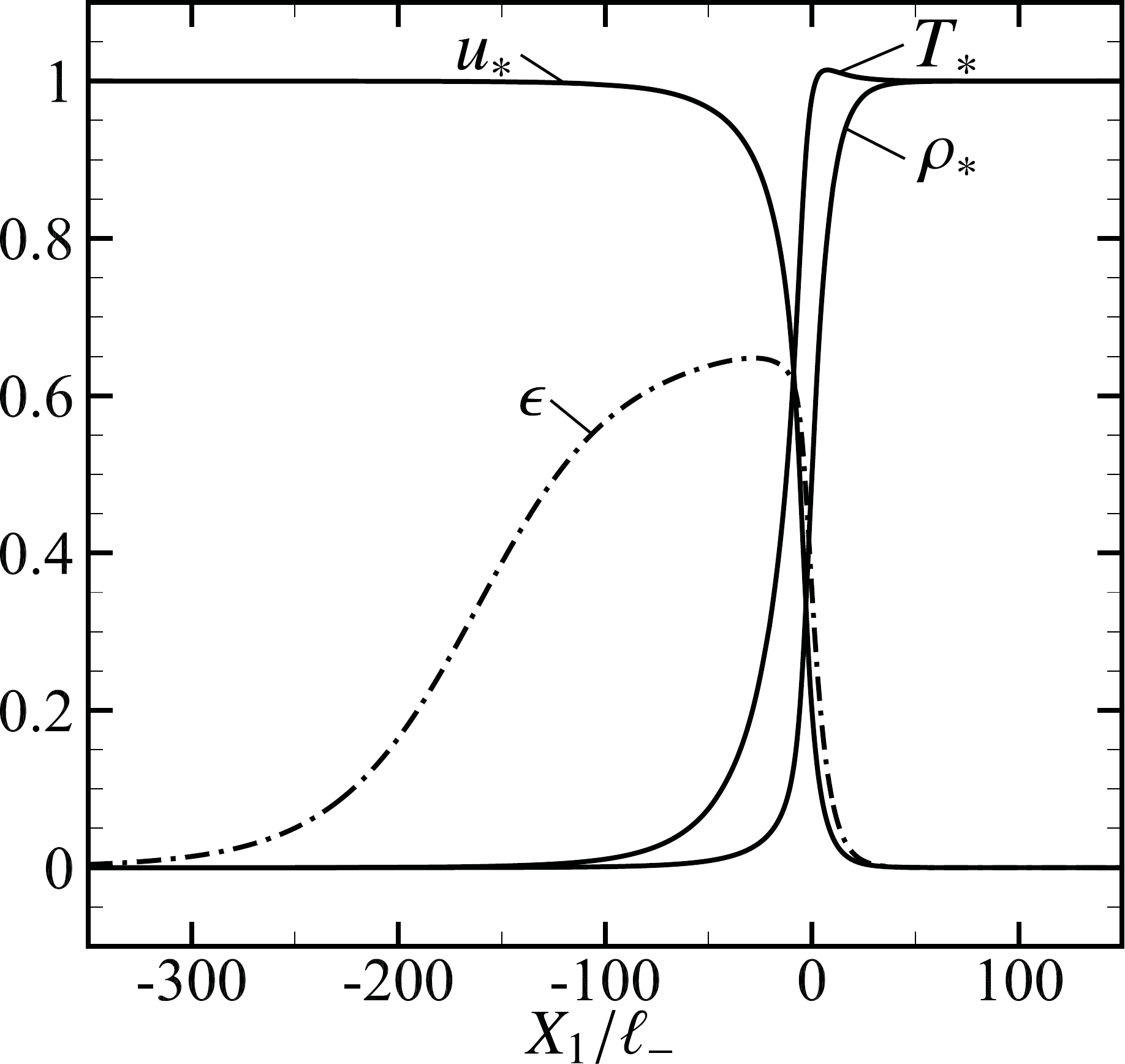} \\
(c) $M_-=20$ & (d) $M_-=40$
\end{tabular}
\caption{Structure of the shock wave for typical Mach numbers $M_-$ 
in the case of $\mathrm{Pr}=2/3$. (a) $M_-=2$, (b) $M_-=5$, (c) $M_-=20$,
and (d) $M_-=40$. Here, $\ell_-$ is the mean free path of molecules 
in the equilibrium state at rest with density $\rho_-$ and temperature $T_-$.
In each panel, solid lines indicate 
$\rho_*=(\rho-\rho_-)/(\rho_+-\rho_-)$,
$u_*=(v_1-u_+)/(u_--u_+)$, and $T_*=(T-T_-)/(T_+-T_-)$ respectively,
while a dash-dotted line indicates $\epsilon$.}
\label{fig:shock}
\end{figure}

Consider a normal shock wave standing at rest against a uniform flow of a monatomic gas.
Let $X_1$ be the spatial coordinate along the flow direction whose origin is located at the center of the shock.%
\footnote{The center of the shock is defined by the position where the density $\rho$ takes the average of the densities at the upstream and downstream infinities.}
Let $\rho_-$, $T_-$, and $(u_-,0,0)$  be the density, temperature, and flow velocity of the  uniform equilibrium state at upstream infinity, while $\rho_+$, $T_+$, and $(u_+,0,0)$ be those at downstream infinity.
Then, 
under the spatially uniform and the null flow-velocity assumption
in the $X_2$- and $X_3$-directions,
the problem is formulated as
\begin{subequations}\begin{align}
& \xi_1\frac{\partial f}{\partial X_1}=Q_\mathrm{ES}(f), \quad -\infty<X_1<\infty, \label{eq:Shock}\\
\intertext{under the upstream and the downstream condition}
&  f(X_1,\bm\xi)\to \frac{\rho_\pm}{(2\pi RT_\pm)^{3/2}}\exp\Big(-\frac{(\xi_1-u_\pm)^2+\xi_2^2+\xi_3^2}{2RT_\pm}\Big),\quad\mbox{as }\ X_1\to\pm\infty,
\end{align}
\end{subequations}
where the upstream and the downstream quantities are related 
by the Rankine--Hugoniot conditions \cite{LR01}:
\begin{equation}
 \frac{\rho_-}{\rho_+}
=\frac{u_+}{u_-}
=1-\frac{2}{\gamma+1}\frac{M_-^2-1}{M_-^2},\quad
\frac{T_+}{T_-}
=1+\frac{2(\gamma-1)}{(\gamma+1)^2}
   \frac{(\gamma M_-^2+1)(M_-^2-1)}{M_-^2}.
\label{RHcond}
\end{equation}
Here $\gamma=5/3$ and $M_-(\ge1)$ is the Mach number of the flow at upstream infinity,
i.e., $M_-=u_-/\sqrt{\gamma RT_-}$.

The problem can be reduced to that of the marginal VDFs \cite{C65}.
After the reduction, 
it is solved numerically by the standard iterative finite-difference method.
The numerical accuracy is carefully checked 
enough to assure the main conclusions drawn from the present computations.
Here we omit the details of the numerical method itself, grid resolution, etc.
and simply present the obtained results.

It is readily seen that $\Theta$ is diagonal%
\footnote{Thanks to the assumption made just before \eqref{eq:Shock},
the problem is rotationally invariant around the $X_1$-axis.
The off-diagonal components are thus null.}
in the present coordinates system,
 and its diagonal components are nothing else than the eigenvalues $\lambda_1$, $\lambda_2$, and $\lambda_3$ [or $(1-\epsilon)\lambda$, $(1-\epsilon\theta)\lambda$, and $\lambda$].
Firstly, the computations have been performed 
for $M_-=2$, $5$, $10$, 20, $40$, and $60$
under the setting $\mathrm{Pr}=2/3$.
Figure~\ref{fig:shock} shows the structure of the shock wave in some typical cases.
In all the cases, $\Theta_{11}$ is found to be the largest among the diagonal components.
Because the problem is isotropic in the directions normal to $X_1$-direction,
$\Theta_{22}=\Theta_{33}$ and thus $\theta\equiv 1$.
Then, the condition \eqref{CC} is simplified as
\begin{equation}
 0\le \epsilon <
 \mathcal{F}(1;\nu)=\frac{9}{2}\frac{1+\nu}{(\nu+2)^2},
\end{equation}
or equivalently
\begin{equation}
 0\le \epsilon < \frac{(\mathrm{Pr}-1/2)}{(\mathrm{Pr}-1/3)^2}\mathrm{Pr}\equiv\mathcal{S}(\mathrm{Pr}),
\label{ShPr}
\end{equation}
using the relation $\nu=1-1/\mathrm{Pr}$. 
Note that $\mathcal{S}(\mathrm{Pr})$ monotonically increases from $0$ to $1$ as $\mathrm{Pr}$ increases from $1/2$ to $2/3$.
The obtained profiles of $\epsilon$, which is given by $1-\Theta_{33}/\Theta_{11}$ here, 
inside the shock are also shown by dash-dotted lines in Figure~\ref{fig:shock}.
With the aid of \eqref{ShPr}, the lower bound of the acceptable Prandtl number for each Mach number
is suggested from the figure, the result of which is shown in Table~\ref{tab:shock}.%
\begin{table}\centering
\caption{The maximum $\epsilon$ for various Mach numbers $M_-$ in the case of $\mathrm{Pr}=2/3$. $\mathrm{Pr}_*$ is the lower bound of acceptable Prandtl number suggested 
by $\max\epsilon=\mathcal{S}(\mathrm{Pr}_*)$; see \eqref{ShPr}.}
\begin{tabular}{ccccccc}
\hline
$M_-$ & 2 & 5 & 10 & 20 & 40 & 60 \\
\hline
$\max\epsilon$  & 0.2786 & 0.5506 & 0.6137 & 0.6387 & 0.6478 & 0.6499 \\
$\mathrm{Pr}_*$ & 0.5184 & 0.5454 & 0.5539 & 0.5576 & 0.5590 & 0.5593 \\
$(3/2)\mathrm{Pr}_*$ & 0.7776 & 0.8181 & 0.8309 & 0.8364 & 0.8385 & 0.8390 \\
\hline
\end{tabular}
\label{tab:shock}
\end{table}
\begin{figure}\centering
\includegraphics[scale=0.3]{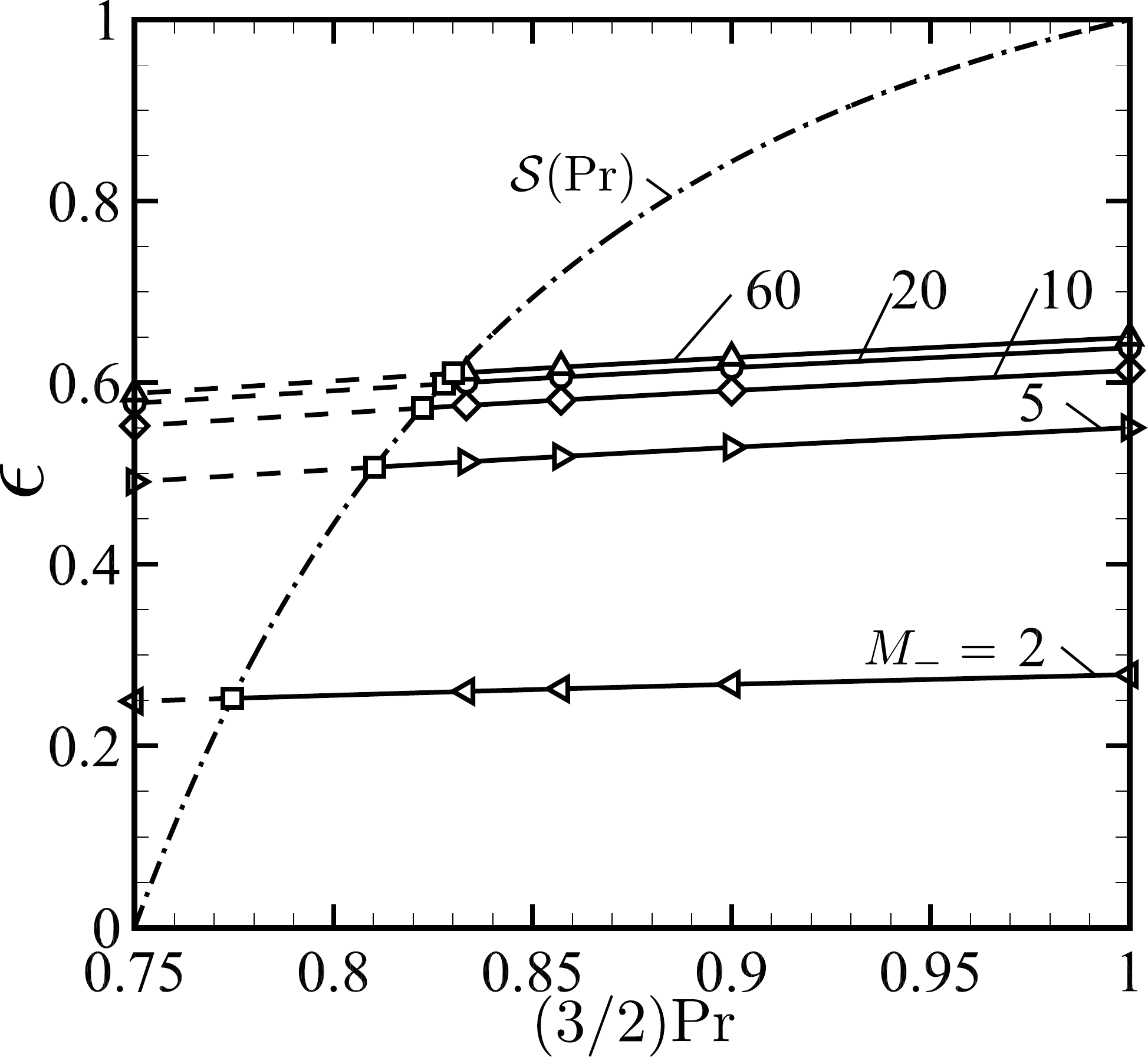}
\caption{The maximum value of $\epsilon$ vs. $(3/2)\mathrm{Pr}$ for various upstream Mach numbers $M_-$.
Symbols indicates the results of computations. The results of common $M_-$ are connected by solid lines in the acceptable range of $\mathrm{Pr}$ and by dashed lines in the range where condition \eqref{CC} is violated.}
\label{fig:shockEps}
\end{figure}
With these results in mind, 
further computations have been carried out for different Prandtl numbers as well.
The maximum values of $\epsilon$ in individual computations 
are plotted against $\mathrm{Pr}$ for each Mach number in Figure~\ref{fig:shockEps}.
It is seen that they depend only weakly on $\mathrm{Pr}$ for every fixed value of $M_-$.
The intersections of solid lines and the curve $\epsilon=\mathcal{S}(\mathrm{Pr})$ 
are indicated by the symbol $\Box$. 
Their abscissas indicate the actual lower bounds of acceptable Prandtl number for the individual $M_-$.
They are all slightly smaller than the values of $\mathrm{Pr}_*$ in Table~\ref{tab:shock} for those $M_-$. 
It is seen that in the present test 
the condition \eqref{CC} [or \eqref{ShPr}] is tolerant;
even for $M_-=60$, $(3/2)\mathrm{Pr}$ can be down to $\simeq 0.83$, i.e., 
a reduction of $\mathrm{Pr}$ by around $17\%$ from $2/3$ is acceptable.

Incidentally, even for $\mathrm{Pr}=1/2$ (or $\nu=-1$), 
where the condition \eqref{ShPr} is violated,
stable numerical computations have been carried out.
It would be partially explained by that
the condition \eqref{cond_T} for $\mathcal{T}$ to be positive definite is reduced to $0\le\epsilon<3/4$
and is never violated in those computations (see Figure~\ref{fig:shockEps}).

\subsection{Couette flow and sliding plate test}

Consider a monatomic rarefied gas in the gap between two parallel plates
which are kept at a common uniform temperature $T_0$. 
The confined gas is in the equilibrium state at rest with the temperature $T_0$ and the density $\rho_0$.
Set the $X_1$-direction to be normal to the plates, with its origin at the middle of the gap. Let the gap width be $L$. At the instance $t=0$, the two plates suddenly start sliding with a constant speed $V$ in the directions opposite to each other: the plate at $X_1=L/2$ moves in the $X_2$-direction, while that at $X_1=-L/2$ in the $-X_2$-direction.
Assuming the diffuse reflection boundary condition on the plates, 
the problem is formulated as
\begin{subequations}
\begin{align}
  &\frac{\partial f}{\partial t}
+\xi_1\frac{\partial f}{\partial X_1}=Q_\mathrm{ES}(f), \quad -L/2<X_1<L/2, \\
\intertext{under the initial and the boundary condition}
  & f(0,X_1,\bm\xi)=\frac{\rho_0}{(2\pi RT_0)^{3/2}}\exp\Big(-\frac{\xi_1^2+\xi_2^2+\xi_3^2}{2RT_0}\Big), \\
  & f(t,\pm L/2,\bm\xi)=\frac{\sigma_\pm}{(2\pi RT_0)^{3/2}}\exp\Big(-\frac{\xi_1^2+(\xi_2\mp V)^2+\xi_3^2}{2RT_0}\Big),\quad \xi_1\lessgtr 0,\\
\intertext{where}
& \sigma_\pm=\sqrt{\frac{2\pi}{RT_0}}\int_{\xi_1\gtrless 0}|\xi_1|f(t,\pm L/2,\bm\xi)d\bm{\xi}.
\end{align}
\end{subequations}
This is the time-dependent Couette flow problem.

As the previous test case, the problem can be reduced to that of the marginal VDFs \cite{C65}.
After the reduction, 
it is solved numerically by the standard finite-difference method implicit in time.
The numerical accuracy is carefully checked 
enough to assure the main conclusions drawn from the present computations.
We again omit all the details of the numerical computations.
In the present test, $\Theta$ is not diagonal in the specified Cartesian coordinates, 
and the three eigenvalues are given by $\Theta_{33}$ and $\lambda_\pm\equiv [\Theta_{11}+\Theta_{22}\pm\sqrt{(\Theta_{11}-\Theta_{22})^2+4\Theta_{12}^2}]/2$.

The computations have been carried out for various values of $V$, the reference Knudsen number, and the Prandtl number.
In the present test, $\epsilon$ is found to be always $1-\lambda_-/\lambda_+$ 
and to take its maximum value on the plates 
at the initial stage of the time evolution.
Consequently, that value is found to depend only on the plate speed, 
neither on the Knudsen number nor the Prandtl number, 
because at the initial stage the collision integral is not effective.
These observations imply that
the most severe situation is identical to the initial stage of the Rayleigh problem \cite{B67,S64}
and is represented by the combination of the initial data for incident molecules and the boundary condition for reflected molecules on the plates.
That is,
the most severe test can be conducted by the following VDF:
\begin{equation}
f(\bm\xi)=\Biggl\{
\begin{array}{ll}
\displaystyle
\frac{\rho_0}{(2\pi RT_0)^{3/2}}
\exp\Big(-\frac{\xi_1^2+(\xi_2-V)^2+\xi_3^2}{2RT_0}\Big), & \xi_1<0,\\
\displaystyle
\frac{\rho_0}{(2\pi RT_0)^{3/2}}
\exp\Big(-\frac{\xi_1^2+\xi_2^2+\xi_3^2}{2RT_0}\Big), & \xi_1>0.
\end{array}
\Biggr .
\label{FMF}
\end{equation}
%
Note that this is identical to the steady free molecular solution of the Couette flow
with one plate sliding with speed $V$ and the other plate at rest.
The above reduction of test, which we call a sliding plate test, allows us to assess the condition of our interest in detail.

\begin{figure}\centering
\includegraphics[scale=0.3]{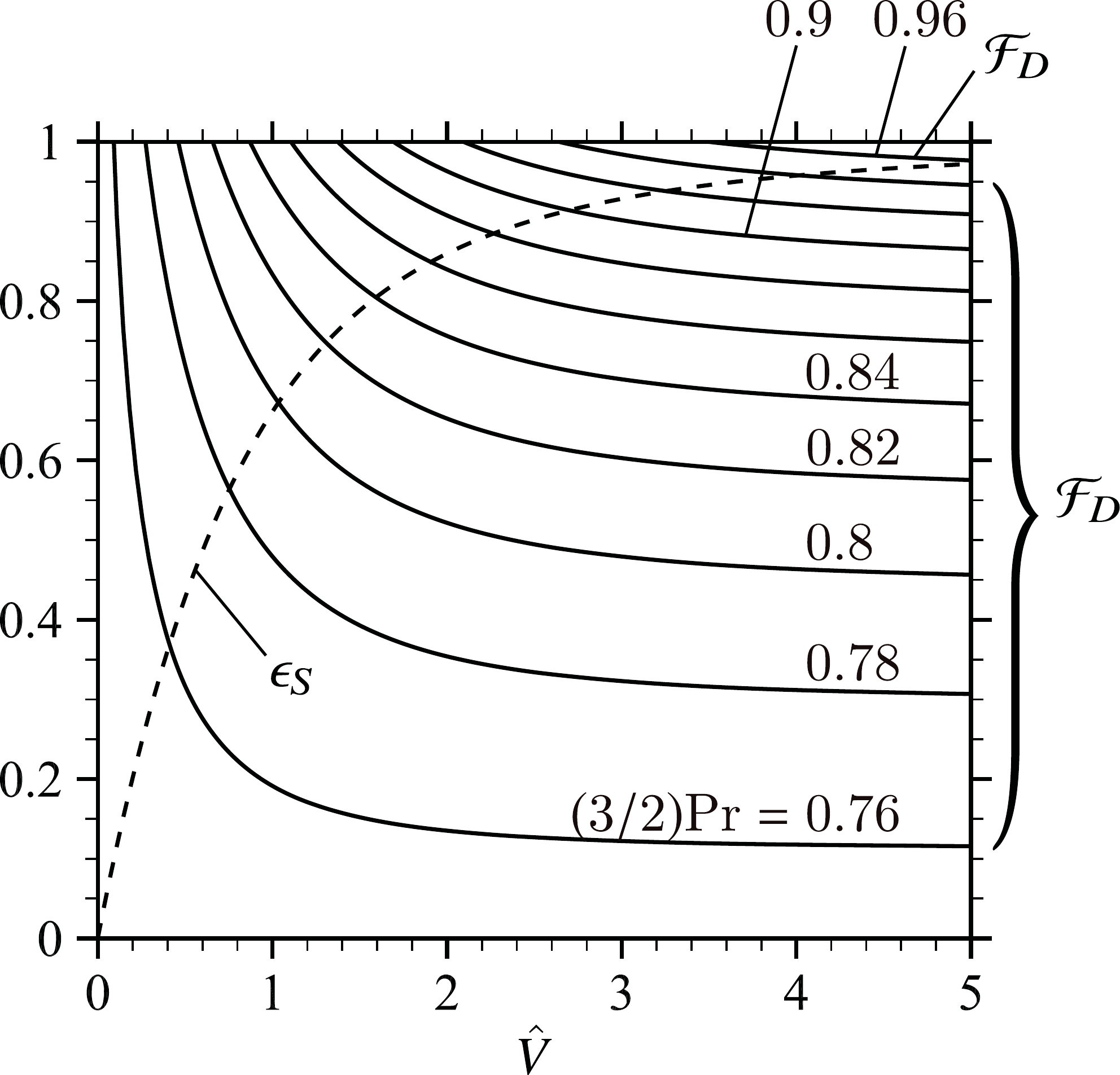}
\caption{The functions 
$\mathcal{F}_D$ and $\epsilon_S$ in the criterion \eqref{cond_D_epsthe}. 
The solid lines indicate $\mathcal{F}_D$ for $(3/2)\mathrm{Pr} = 0.76, 0.78, 0.8,\dots,0.96$,
while the dashed line indicates $\epsilon_S$.}
\label{fig:D_epsthe}
\end{figure}

For the above VDF \eqref{FMF},  $\epsilon$ and $\theta$ are readily obtained as
\begin{equation}
\epsilon=\epsilon_S(\hat{V})\displaystyle
\equiv\frac{\displaystyle\frac{\hat{V}^2}{2}\sqrt{1+\frac{16}{\pi \hat{V}^2}}}{1+\displaystyle\frac{\hat{V}^2}{4}(\sqrt{1+\displaystyle\frac{16}{\pi \hat{V}^2}}+1)},\quad
\theta=\theta_S(\hat{V})\displaystyle
\equiv\frac12\Big(1+\frac{1}{\sqrt{1+\displaystyle\frac{16}{\pi\hat{V}^2}}}\Big),
\label{epsthe}
\end{equation}
where $\hat{V}=V/\sqrt{2RT_0}$. Note that (a) $\theta>1/2$ from the above, while $0<\theta_a\le1/2$ in the range $-1\le\nu<-1/2$ or $1/2\le\mathrm{Pr}<2/3$. It is also found that $0\le\epsilon_S(\hat{V})<1$. 
Hence, plugging \eqref{epsthe} into \eqref{CC} yields the following criterion
for the H theorem to remain valid in the present test:
%
\begin{equation}
0\le \epsilon_S(\hat{V}) <
 \frac{9(2-1/\mathrm{Pr})[\theta_S(\hat{V})^2-\theta_S(\hat{V})+1]}{[1+\theta_S(\hat{V})]
 F(\theta_S(\hat{V});1-1/\mathrm{Pr})}
 \equiv \mathcal{F}_D(\hat{V};\mathrm{Pr}),
\label{cond_D_epsthe}
\end{equation}
%
where the relation $\nu=1-1/\mathrm{Pr}$ has been used to convert $\nu$ to $\mathrm{Pr}$.
The function $\mathcal{F}_D$ is plotted by solid lines for various values of $\mathrm{Pr}$ 
in Figure~\ref{fig:D_epsthe}, together with $\epsilon_S$ by a dashed line.
It is seen from the figure that
, by the 20 \% (resp. 16 \%) reduction of Prandtl number from $2/3$, 
the condition \eqref{cond_D_epsthe} is satisfied for the sliding speed $\hat{V}\lesssim 1.03$ (resp. 1.59).

Incidentally, in the time-dependent Couette flow test, 
even when the condition \eqref{CC} is violated,
numerical computations have been carried out stably and consistently, 
as far as the condition \eqref{cond_T} is satisfied.
This observation agrees with the shock wave test for $\mathrm{Pr}=1/2$.

\begin{figure}
\centering
\begin{tabular}{cc}
\includegraphics[scale=0.3]{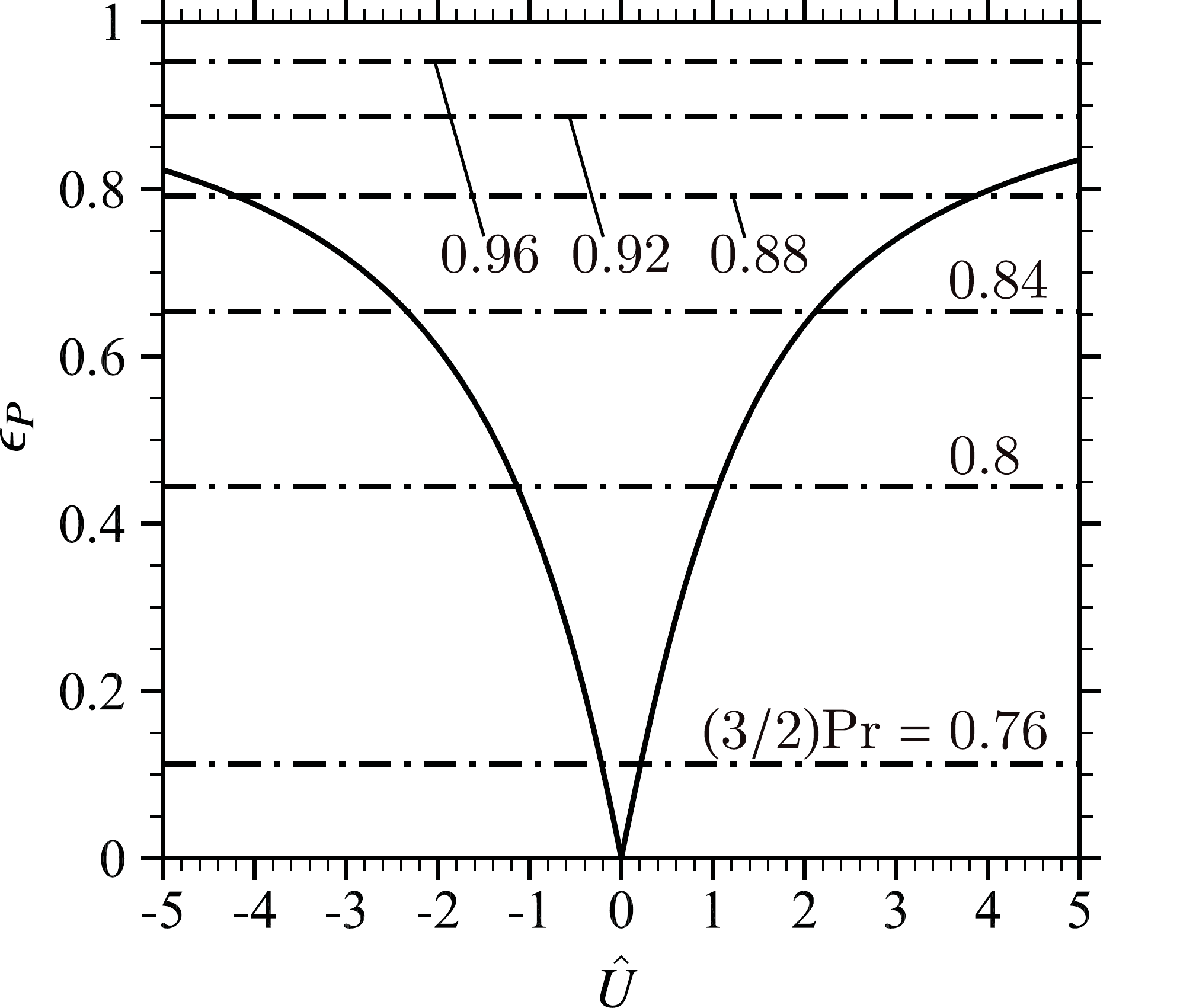} &
\includegraphics[scale=0.3]{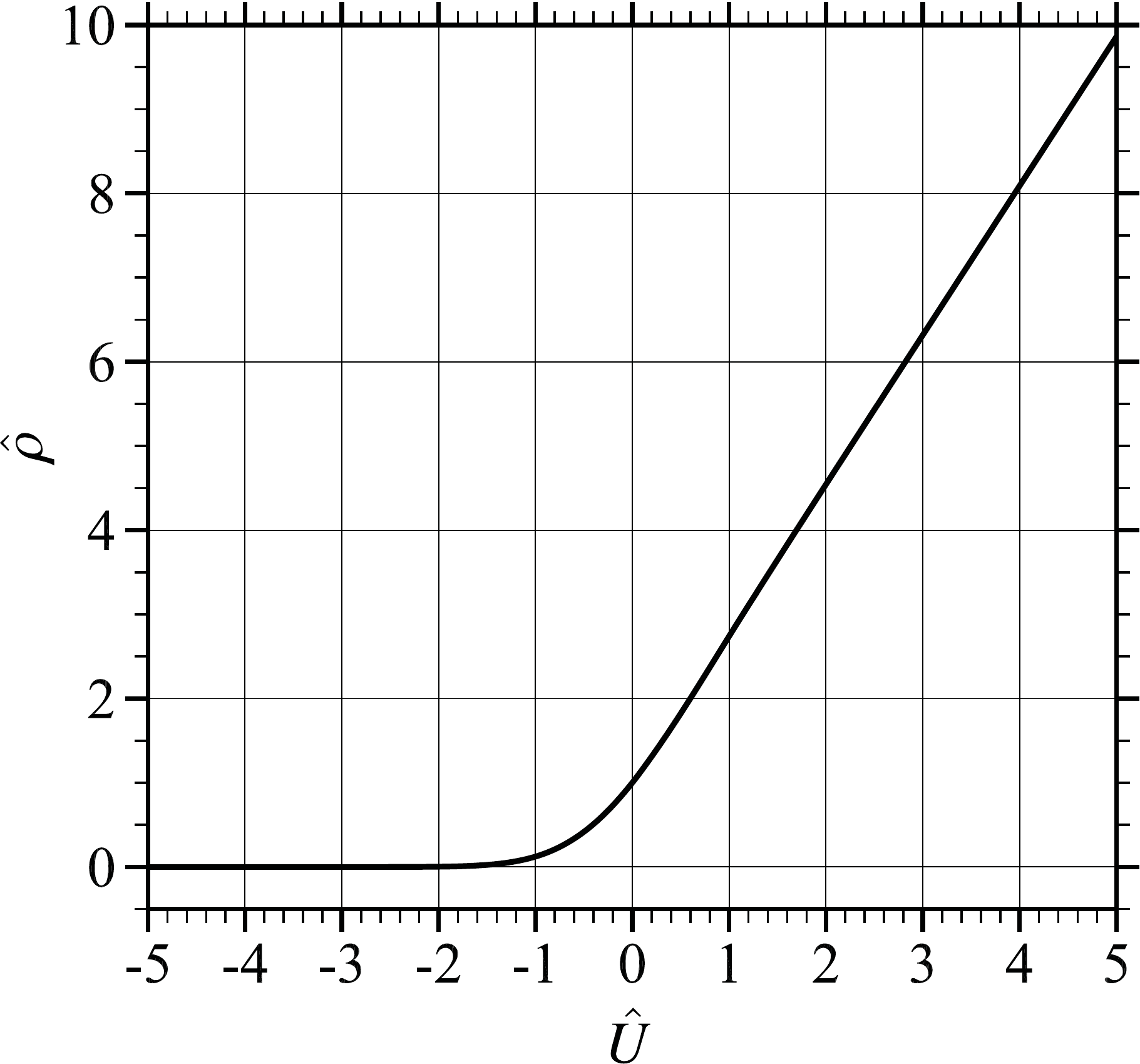}\\
(a) & (b)
\end{tabular}
\caption{The function $\epsilon_P$ and the dimensionless density $\hat{\rho}$ in the range $-5<\hat{U}<5$. 
(a) $\epsilon_P$, (b) $\hat{\rho}$. In (a), the values of $\mathcal{S}(\mathrm{Pr})$ for $(3/2)\mathrm{Pr}=0.76,0.8,0.84,\dots,0.96$ are also indicated 
by dash-dotted lines for reference.}
\label{fig:rhoeps}
\end{figure}

\subsection{Pushing or pulling plate test}
Motivated by the sliding plate test, 
we consider another severe test by the following VDF:
\begin{equation}
f(\bm\xi)=\Biggl\{
\begin{array}{ll}
\displaystyle
\frac{\rho_0\hat{\sigma}}{(2\pi RT_0)^{3/2}}
\exp\Big(-\frac{(\xi_1-U)^2+\xi_2^2+\xi_3^2}{2RT_0}\Big), & \xi_1> U,\\
\displaystyle
\frac{\rho_0}{(2\pi RT_0)^{3/2}}
\exp\Big(-\frac{\xi_1^2+\xi_2^2+\xi_3^2}{2RT_0}\Big), & \xi_1< U,
\end{array}
\Biggr .
\label{VM}
\end{equation}
where 
\begin{align*}
& \hat{\sigma}(\hat{U})
=\frac{1}{\rho_0}
 \sqrt{\frac{2\pi}{RT_0}}\int_{\xi_1< U}|\xi_1-U|f d\bm{\xi}
=\exp(-\hat{U}^2)+\sqrt{\pi}\hat{U}[1+\mathrm{erf}(\hat{U})], \\
& \hat{U}=U/\sqrt{2RT_0}.
\end{align*}
The above VDF mimics the gas on the plate at the initial instance of a sudden plate motion 
in the normal direction with a constant speed $|U|$,
by supposing that, before starting the plate motion,
the gas is in the equilibrium state at rest with density $\rho_0$ and temperature $T_0$.
Here,
$U>0$ implies the pushing plate while $U<0$ implies the pulling plate.

In this case, $\Theta$ is diagonal with $\Theta_{22}=\Theta_{33}$ (thus $\theta=1$) as in the shock wave test, and the diagonal components are given by
\begin{equation}
 \Theta_{11}/(RT_0)
=1+({\hat{U}}/{\sqrt{\pi}})\hat{\sigma}(\hat{U})
 /\hat{\rho}(\hat{U}),
\quad
\Theta_{22}=\Theta_{33}=RT_0,
\end{equation}
where $\hat{U}=U/\sqrt{2RT_0}$ and 
$\hat{\rho}(\hat{U})=(1/2)[\hat{\sigma}(\hat{U})+1+\mathrm{erf}(\hat{U})]$
is the gas density divided by $\rho_0$.
Note that $\hat{\sigma}>0$ because of its definition. 
Accordingly, $\Theta_{11}\gtrless RT_0$ for $\hat{U}\gtrless 0$
and $\epsilon$ is written as
\begin{equation}
\epsilon=1-(RT_0/\Theta_{11})^{\pm1},\quad \hat{U}\gtrless 0,
\label{pushpull}
\end{equation}
in the present test.
Since $\theta=1$, the condition \eqref{CC} takes the same form as \eqref{ShPr}.
Substitution of \eqref{pushpull} into \eqref{ShPr} eventually
leads to the following condition:
\begin{equation}
 0\le\epsilon_P(\hat{U})<\mathcal{S}(\mathrm{Pr}),
\end{equation}
where
\begin{equation}
 \epsilon_P(\hat{U})= 1-\Big(1+({\hat{U}}/{\sqrt{\pi}})
 \hat{\sigma}(\hat{U})/\hat{\rho}(\hat{U})
 \Big)^{\mp 1}, \quad \hat{U}\gtrless 0.
\end{equation}
Figure~\ref{fig:rhoeps} shows the functions $\epsilon_P$  
and $\hat{\rho}$ respectively in (a) and (b), 
where the values of $\mathcal{S}(\mathrm{Pr})$ are also shown 
for reference in (a).
The present test assesses the robustness of the desired entropic property for small Prandtl numbers against
the anisotropy caused by strong compression (or expansion).
In this sense, the type of test is similar to that 
in the shock wave case.
Nevertheless, the observed robustness looks quite different.
One possible explanation is the compression rate
measured by the density. 
In the shock wave test, the ratio of density $\rho_+/\rho_-$
 between the upstream and the downstream infinity 
 is at most four even in the limit $M_-\to\infty$, see \eqref{RHcond}.
But here, the ratio of the density to its initial value $\hat{\rho}$
or its inverse is beyond this value already around $\hat{U}=1.70$ or $-0.73$. 
The present test is thus much harder than the shock wave test.

\section{Conclusions}\label{sec:conclusion}

In the present paper, we have investigated the criterion for the H theorem
to hold even for the ES model with the Prandtl number below $2/3$.
The derived condition that covers the range $1/2\le\mathrm{Pr}<2/3$
shows that indeed there is a room for the H theorem to remain valid.
The practical tolerance of the condition is assessed 
numerically by a couple of tests, i.e., 
the strong shock wave and the time-dependent Couette flow problem,
and then further assessed by the sliding and the pushing/pulling plate problem.
The condition is tolerant in the shock wave test
enough to accept about $18\%$ reduction of $\mathrm{Pr}$ below $2/3$
even for the strong shock wave with $M_-=5$,
while it is rather restrictive in the Couette flow, the sliding plate, and the pushing/pulling plate test 
in the sense that the acceptable reduction of $\mathrm{Pr}$ of the same level is achieved only within $1.4$ for both $|\hat{U}|$ and $\hat{V}$.
The latter three tests show that 
the collisionless gas state caused by the sudden motion of boundary 
can be a practical source of restriction on the entropic property
of the ES model for small Prandtl numbers below $2/3$.

\appendix
\section{A brief review of the H theorem in Ref.~\cite{ATPP00}}
\label{sec:app}

Proposition 2.1 (the H theorem) of Ref.~\cite{ATPP00},
together with the outline of its proof, 
is shown here in terms of the notation in the present paper.

Let 
\begin{equation}
S(\rho,\bm{v},\mathcal{T})=\min_{g\in\mathcal{X}}\langle H(g)\rangle,
\label{eq:minimizer}
\end{equation}
where $H(f)=f\ln f$, $\langle\bullet\rangle=\int\bullet\, d\bm\xi$, and $\mathcal{X}(\rho,\bm{v},\mathcal{T})$ is the set defined by
$\mathcal{X}=\{g\ge0, (1+\xi_i^2)g\in L^1(\mathbb{R}^3),
\langle g\rangle=\rho, \langle\xi_i g\rangle=\rho v_i,
\langle \xi_i\xi_j g \rangle=\rho v_iv_j+\rho\mathcal{T}_{ij}\}$.
Then, the following proposition is proved in Ref.~\cite{ATPP00}.

\medskip
\noindent\textbf{Proposition:}\ 
For symmetric positive definite tensor $\Theta$
and $-1/2\le\nu<1$ we have
\begin{enumerate}
\item[(a)] the tensor $\mathcal{T}$ defined in \eqref{eq:CalT} 
is symmetric positive definite and 
the set $\mathcal{X}(\rho,\bm{v},\mathcal{T})$ is not empty,
\item[(b)] the unique minimizer in \eqref{eq:minimizer}
is the Gaussian $\mathcal{G}[f]$ defined in \eqref{eq:Gauss},
\item[(c)] the entropy of the Gaussian $\mathcal{G}[f]$ satisfies
$\langle H(\mathcal{G})\rangle
=S(\rho,\bm{v},\mathcal{T})
\leq S(\rho,\bm{v},\Theta)
\leq \langle H({f})\rangle$,
\item[(d)] consequently the entropy inequalities
\begin{align}
&\int Q_\mathrm{ES}(f)\ln f d\bm{\xi}\le0,\\
&\frac{\partial}{\partial t}\int H(f)d\bm{\xi}
+\frac{\partial}{\partial X_i}\int \xi_i H(f)d\bm{\xi}\le 0,
\end{align}
hold.
\item[(e)] the equality 
$\langle H(\mathcal{G})\rangle=\langle H({f})\rangle$
implies $f=\displaystyle\frac{\rho}{(2\pi RT)^{3/2}}\exp(-\frac{(\bm\xi-\bm{v})^2}{2RT})$.
\end{enumerate}

\medskip
The outline of proof of (a)--(e) given in Ref.~\cite{ATPP00} is as follows.
\begin{enumerate}
\item[On (a):]
The first part is assured by straightforward manipulations 
in terms of the eigenvalues of $\Theta$ 
(i.e., $\lambda_1$, $\lambda_2$, and $\lambda_3$) for $-1/2\le\nu\le 1$.
Then second part is valid for such $\mathcal{T}$, 
because $\mathcal{G}[f]$ belongs to that set.
\item[On (b):]
The proof relies on the definition of $H$ and its convexity.
\item[On (c):]
The equality $\langle H(\mathcal{G})\rangle=S(\rho,\bm{v},\mathcal{T})$
is nothing else than (b). 
The inequality $S(\rho,\bm{v},\mathcal{T})\le S(\rho,\bm{v},\Theta)$
is first reduced to $\det\mathcal{T}\ge\det\Theta$ by using (b).
Then, the reduced inequality is validated by an extension of the Brunn--Minkowsky inequality
in the range $-1/2\le\nu\le 1$.
The inequality $S(\rho,\bm{v},\Theta)\le\langle H(f)\rangle$
is a consequence of the definition of $S$.
\item[On (d):]
This is an immediate consequence of (c) and the convexity of $H$.
\item[On (e):]
Because of (d), 
the equality $\langle H(\mathcal{G})\rangle=\langle H({f})\rangle$
implies $S(\rho,\bm{v},\Theta)=\langle H(f)\rangle$,
which in turn implies that $f$ is the Gaussian with $\mathcal{T}$ being replaced by $\Theta$. It also implies 
$S(\rho,\bm{v},\mathcal{T})=S(\rho,\bm{v},\Theta)$,
which leads to $\det\mathcal{T}=\det\Theta$.
Then, direct calculations of this equality 
show that $\Theta$ is a scalar multiple of the identity matrix
and thus $f$ is the Maxwellian given in (e). 
\end{enumerate}

\medskip
As is clear from the above outline, 
the clue steps for the extension of the validity range of the H theorem
are to show that (i) $\mathcal{T}$ is positive definite
and (ii) $\det\mathcal{T}\ge\det\Theta$, 
including that the equality in (ii) holds only when $f$ is the Maxwellian.
The discussions in Sec.~\ref{sec:Derivation} of the present paper
focus on the above (i) and (ii).



\nocite{*}
\bibliographystyle{aipnum-cp}%

\end{document}